\documentclass{elsart}
\usepackage{amsfonts}
\usepackage{amsmath}
\usepackage{rotating}
\usepackage{epsfig}
\usepackage{epsf}

\input{tcilatex}

\begin{document}

\begin{frontmatter}

\title{Symbolic Dynamics in a Matching Labor Market Model}

\author{Diana A. Mendes*, Vivaldo M. Mendes**, J. Sousa Ramos***}

\address{*Corresponding author: Department of Quantitative Methods and UNIDE, IBS - ISCTE Business School, ISCTE, Av. Forcas Armadas, 1649-026 Lisbon, Portugal. diana.mendes@iscte.pt. 
 **Department of Economics and UNIDE, ISCTE, Lisbon, Portugal. vivaldo.mendes@iscte.pt., 
***Department of Mathematics, IST, Technical University of Lisbon,  Portugal. sramos@math.ist.utl.pt. }

\begin{abstract}
In this paper we apply the techniques of symbolic dynamics to the analysis of a 
labor market which shows large volatility in employment flows. In a recent paper, 
Bhattacharya and Bunzel \cite{BB} have found that the discrete time
 version of the Pissarides-Mortensen matching model can easily lead to chaotic
 dynamics under standard sets of parameter values. To conclude about the 
existence of chaotic dynamics in the numerical examples presented in the paper,
 the Li-Yorke theorem or the Mitra sufficient condition were applied which seems
 questionable because they may lead to misleading conclusions. Moreover, in a 
more recent version of the paper, Bhattacharya and Bunzel \cite{BB1} 
 present new results in which chaos is completely removed from the dynamics 
of the model. Our paper explores the matching model so interestingly developed
 by the authors with the following objectives in mind: (i) to show that chaotic dynamics
 may still be present in the model for standard parameter values; (ii) to clarify some
 open questions raised by the authors in \cite{BB}, by providing a rigorous
 proof of the existence of chaotic dynamics in the model through the computation
 of topological entropy in a symbolic dynamics setting.
\end{abstract}


\end{frontmatter}

\section*{Introduction}

In a recent paper, Levin \cite{Levin} discusses the set of results presented
over the last decade by various prominent physicists which led to the
conclusion that black holes seem to be susceptible to chaos. Levin argues
that the most realistic description available of a spinning pair of black
holes is chaotic motion, and goes on to complain that in physics and
cosmology \textquotedblright chaos has not received the attention it
deserves in part because the systems studied have been highly
idealized\textquotedblright . In contrast, in economics we have the
interesting fact that even some of the most simple and highly idealized
models describing modern economies can easily lead to chaotic dynamics.%
\footnote{%
See, e.g., \cite{Saari}, \cite{Nishimura}, \cite{Benhabib}, \cite{Grandmont}
and \cite{Boldrin}. The potential for very complex behavior significantly
increases if models become somewhat less reductionist, e.g., if heterogenous
agents and different learning processes are also taken into account, \cite%
{Grandmont2}, \cite{Brock}.}

In this paper we apply the techniques of symbolic dynamics to the analysis
of a labor market which shows in almost developed economies large volatility
in employment flows. The possibility that chaotic dynamics may arise in
modern labor markets had been totally strange to economics until recently,
at least as far as we are aware of. In an interesting paper, Bhattacharya
and Bunzel \cite{BB} have found that the discrete time version of the
Pissarides-Mortensen matching model, as formulated in Ljungqvist and Sargent 
\cite{Ljungk}, can easily lead to chaotic dynamics under standard sets of
parameter values. However, in order to conclude about the existence of
chaotic dynamics in the three numerical examples presented in the paper, the
authors apply the Li-Yorke theorem or the Mitra sufficient condition which
should not be generally applied to all specific simulations because they may
lead to misleading conclusions. Moreover, in a more recent version of the
paper, Bhattacharya and Bunzel \cite{BB1} present new results in which chaos
is completely removed from the dynamics of the model. This paper explores
the matching model so interestingly developed by Bhattacharya and Bunzel
with the following objectives in mind: (i) to show that chaotic dynamics may
still be present in the model for standard parameter values, for high values
of the measure of labor tightness (that is the ratio of vacancies to the
number of workers looking for jobs) which can occur in economic booms; (ii)
to clarify some open questions raised by the authors in their first paper,
by providing a rigorous proof of the existence of chaotic dynamics in the
model through the computation of topological entropy in a symbolic dynamics
setting.

Therefore, if one is studying whether there are chaotic dynamics or not
under certain ranges of parameters values, we suggest that a bifurcation
diagram, the variation of the Lyapunov exponent, the existence of a periodic
orbit of period not equal to a power of two and positive topological entropy
are some techniques that can give a clear answer to this problem.

\section{The Matching Model}

Why would the labor market in most of the developed economies behave in such
a volatile way as the evidence points out? One possibility, and usually the
most favoured one in the dominant view of economics, is that the economy\
has an inherently linear structure and is hit by permanent exogenous
shocks.\ As these shocks are entirely unpredictable, they render the
dynamics and the cycles hardly predictable and controllable. Another more
recent view, which should also be considered for discussion because it seems
consistent and realistic, is based on the possibility that the economy has a
structure that is nonlinear and the cycles are an endogenous manifestation
of this characteristic, either with or without external shocks added to the
structure. In what follows, a very simple and fully deterministic model will
be presented that is capable of generating such type of volatility with
standard parameter sets and no noise.

Let us assume that in every period of time there are large flows of workers
moving into and out of employment: a certain number of job vacancies is
posted by firms $(v_{t})$ and there is a total measure of workers looking
for jobs $(u_{t})$. When a worker and a firm reach an agreement there is a
successful match, and the total number of these matches is given by the
aggregate matching function 
\begin{equation}
M\left( u_{t},v_{t}\right) =Au_{t}^{\alpha }v_{t}^{1-\alpha }A>0,\;\alpha
\in \left( 0,1\right) .  \label{EqM(u,v)}
\end{equation}%
The intuition behind (\ref{EqM(u,v)}) is very simple: the higher is $u,$ the
easier it will be for firms to get a worker with the desired qualifications;
and the higher is the level of vacancies posted by firms $v$, the higher is
the probability that a worker will find an appropriate job. For simplicity
we will assume $A$ as a constant. However, a more adequate treatment would
consist of treating $A$ as a variable dependent on the level of public
provision of information by public agencies with the objective of increasing
the number of successful matches.

The measure of labor tightness is given by the ratio $\theta _{t}\equiv
v_{t}/u_{t}.$ Then, the probability of a vacancy being filled at $t$ is
given by%
\begin{equation*}
q\left( \theta _{t}\right) \equiv \frac{M\left( u_{t},v_{t}\right) }{v_{t}}%
=A\theta _{t}^{-\alpha },\ 0<A<1.
\end{equation*}

Let $n_{t+1}$ be the total number of employed workers at the beginning of $%
t+1$ and let $s$ be defined as the probability of a match being dissolved at 
$t$. Therefore we have 
\begin{equation*}
n_{t+1}=\left( 1-s\right) n_{t}+q\left( \theta _{t}\right) v_{t},
\end{equation*}
where $\theta _{t}\equiv v_{t}/u_{t}=v_{t}/(1-n_{t}).$ Notice that $\left(
1-s\right) n_{t}$ gives the number of undissolved matches prevailing at $t$
and passed on to $t+1$, while $q\left( \theta _{t}\right) v_{t}$ represents
the number of new matches formed at $t$ with the available number of
unemployed workers and vacancies.

As shown in \cite{Ljungk}, the model can be solved for the decentralized
outcome of a Nash bargaining game between workers and firms but to keep the
model as close as possible to the presentation in \cite{BB} and \cite{BB1}
we should focus upon the central planner solution to the matching model. The
objective function of the central planner is given by 
\begin{equation*}
U(n,v)=\phi n_{t}+z\left( 1-n_{t}\right) -cv_{t}
\end{equation*}%
where $\phi $, $z$ and $c$ are parameters that represent, respectively, the
productivity of each worker, the lost value of leisure due to labor effort,
and the cost that firms incur per vacancy placed in the market.\footnote{%
Notice the trade--off between vacancies and unemployment in this objective
function. The first right hand term represents the benefits to society from
successful matches (working), while the last two give the leisure costs and
the costs associated with posting vacancies.} Therefore, the planner chooses 
$v_{t}$ and the next period's employment level, $n_{t+1},$ by solving the
following dynamic optimization problem%
\begin{equation*}
\max_{v_{t},n_{t+1}}\sum_{t=0}^{\infty }\beta ^{t}\left[ \phi n_{t}+z\left(
1-n_{t}\right) -cv_{t}\right]
\end{equation*}%
subject to%
\begin{equation*}
n_{t+1}=\left( 1-s\right) n_{t}+q\left( \frac{v_{t}}{1-n_{t}}\right) v_{t},
\end{equation*}%
where $\beta $ is the time discount rate and an initial condition $n_{0}$ is
given. The Lagrangian can be written as 
\begin{equation*}
L=\sum_{t=0}^{\infty }\left\{ \beta ^{t}\left[ \phi n_{t}+z\left(
1-n_{t}\right) -cv_{t}\right] +\lambda _{t}\left[ \left( 1-s\right)
n_{t}+q\left( \frac{v_{t}}{1-n_{t}}\right) v-n_{t+1}\right] \right\} .
\end{equation*}%
The first order conditions (FOC), for an interior solution, are given by%
\begin{gather*}
\frac{\partial L}{\partial v_{t}}=-\beta ^{t}c+\lambda _{t}\left[ q^{\prime
}\left( \theta _{t}\right) \theta _{t}+q\left( \theta _{t}\right) \right] =0
\\
\frac{\partial L}{\partial n_{t+1}}=-\lambda _{t}+\beta ^{t+1}\left( \phi
-z\right) +\lambda _{t+1}\left[ \left( 1-s\right) +q^{\prime }\left( \theta
_{t+1}\right) \theta _{t+1}^{2}\right] =0.
\end{gather*}

The very interesting point in \cite{BB} and \cite{BB1} was the manipulation
of these FOC to arrive at a reduced equation that can lead to chaotic
dynamics. From the first FOC we get $\lambda _{t}=\frac{\beta ^{t}c}{%
q^{\prime }\left( \theta _{t}\right) \theta _{t}+q\left( \theta _{t}\right) }
$ and substituting this and the corresponding expression for $\lambda _{t+1}$
into the second FOC we obtain 
\begin{equation}
a\theta _{t+1}^{\alpha }-b\theta _{t+1}=\theta _{t}^{\alpha }-d,\text{ \ }%
\alpha \in \left( 0,1\right)  \label{EqPricipal}
\end{equation}%
with the following parameter definitions and restrictions%
\begin{equation}
a\equiv \beta \left( 1-s\right) \in \left( 0,1\right) ,\ 1>b\equiv A\alpha
\beta >0,\ d\equiv \left( A/c\right) \left( 1-\alpha \right) \beta \left(
\phi -z\right) >0.  \label{EqParam}
\end{equation}

Equation (\ref{EqPricipal}) gives the law of motion for the index of labor
market tightness in the economy under the planner's solution. In other
words, given an initial condition $\theta _{0},$ equation (\ref{EqPricipal})
completely characterizes the trajectory of $\theta $ and the whole economy$.$
So, the backward dynamics of this model can be characterized by the
four-parameter one-dimensional family of maps $g:\left[ 0,g_{\max }\right]
\rightarrow \left[ 0,g_{\max }\right] ,$ where%
\begin{equation}
g\left( \theta \right) =\left( a\theta ^{\alpha }-b\theta +d\right) ^{\frac{1%
}{\alpha }},  \label{EqMapa}
\end{equation}%
with the parameter restrictions in (\ref{EqParam}) and $\theta _{\max
}=\left( \frac{\alpha a}{b}\right) ^{\frac{1}{1-\alpha }}$ where $g_{\max }$
is implicitly defined as the lowest positive root of the equation $ag_{\max
}^{\alpha }-bg_{\max }+d=0.$ The first derivative of the map $g$ can be
calculated as 
\begin{equation*}
g^{\prime }\left( \theta \right) =\left( a\theta ^{\alpha }-b\theta
+d\right) ^{\frac{1-\alpha }{\alpha }}\left( a\theta ^{\alpha -1}-\frac{b}{%
\alpha }\right) ,\;\theta \in \left[ 0,g_{\max }\right] ,
\end{equation*}%
which implies that $g$ is unimodal with a unique maximum (critical point) at 
$\theta _{\max }.$ In addition, $g$ has a unique fixed point located to the
right of $\theta _{\max }$ if $g\left( \theta _{\max }\right) >\theta _{\max
}.$ The unique fixed point of $g$ is denoted by $\theta _{\ast }$ and is
implicitly given by $a\theta _{\ast }^{\alpha }-b\theta _{\ast }=\theta
_{\ast }^{\alpha }-d.$ Despite the impossibility of the computation of an
explicit expression for $\theta _{\ast },$ the unicity of this solution is
obvious by considering $\ f_{1}\left( \theta \right) =a\theta _{\ast
}^{\alpha }-b\theta _{\ast }$ and\ $\ f_{2}\left( \theta \right) =\theta
_{\ast }^{\alpha }-d,$where $f_{1}\left( \theta \right) $ is monotonically
decreasing for $\theta $ from $\theta _{\max }$ to $+\infty $ and $%
f_{2}\left( \theta \right) $ is monotonically increasing for $\theta $ from $%
0$ to $+\infty .$ Therefore $f_{1}\left( \theta \right) =f_{2}\left( \theta
\right) $ has a unique solution for $\theta >\theta _{\max }.$

The fixed point is an attractor in the case of backward dynamics if $\left|
g^{\prime }\left( \theta _{\ast }\right) \right| <1$ and in forward dynamics
if $g^{\prime }\left( \theta _{\ast }\right) <-1.$\ For $g^{\prime }\left(
\theta _{\ast }\right) =-1$ a period-doubling bifurcation occurs and the
fixed point changes stability. Since it is not possible to obtain a closed
form expression for $\theta _{\ast },$ this condition cannot be checked in
general but can be checked for each set of parameters separately.

\section{Chaotic Dynamics in the Model}

The first main objective is to show that the matching model leads to chaotic
motion in the backward dynamics and the second purpose is to provide a very
rigorous method to compute topological entropy for uni(multi)-modal maps.
Generally, the search for chaos in one-dimensional maps is mainly based on
the application of the well-known Li-Yorke theorem, but when a period three
fails to be encountered it is necessary to use some other criteria. In what
follows, we use the kneading sequence (the trajectory initiated at the
critical point of the unimodal map) and the associated Markov transition
matrix in order to compute the topological entropy of the map. These are
tools from symbolic dynamics which provide simple and exact methods to
conclude about the existence of chaos in multimodal maps.

In \cite{BB1} the authors conclude that the unique fixed point $\theta
_{\ast }$ of the map (\ref{EqMapa}) is always stable in backward dynamics,
for almost any choice of the parameters values with respect to conditions (%
\ref{EqParam}). We obtain a different result: while for most reasonable sets
of parameter values the fixed point is stable, however there are also other
sets of parameter values that allow for unstable chaotic motion in the model
as we show next.

\begin{prop}
For $\alpha =0.21,$ $\beta =0.955,$ $A=0.99725,$ $\gamma =(\phi -z)/c=1.31,$ 
$s=0.1518,$ we obtain chaotic motion for the nonlinear map presented in (\ref%
{EqMapa}). Moreover, this is true for almost any value of $\alpha $ in the
interval $[0.2,0.22]$ .
\end{prop}

\begin{pf}
The first point which we would like to emphasize is that all parameter
values satisfy the restrictions presented in equation (\ref{EqParam}). That
is%
\begin{eqnarray*}
a &\equiv &\beta \left( 1-s\right) =0.21\in \left( 0,1\right) ,\ 1>b\equiv
A\alpha \beta =0.2>0,\  \\
d &\equiv &A\left( 1-\alpha \right) \beta \gamma =0.9856>0.
\end{eqnarray*}%
This is illustrated in Figure \ref{fig1}, where the unimodal map and the
typical chaotic time series associated to the unstable fixed point $\theta
_{\ast }=3.6793$ are presented. The first derivative of the unimodal map is $%
g^{\prime }\left( \theta _{\ast }\right) =-1.8554$, which shows that the
equilibrium is unstable. By varying the parameter $\alpha $ in the interval $%
[0.2,0.3]$ we obtain the classical period-doubling route to chaos which is
also illustrated in Figure \ref{fig1}. So, for any value of $\alpha $
arising after the $2^{\infty }$ bifurcation point, we encounter chaotic
dynamics. We should point out that in order to obtain the bifurcation
diagram illustrated in Figure \ref{fig1} the parameters $a,b,$ and $d$ have
also to have their values changed, as they are continuously depending on the
values of $\alpha .$ The parameter restrictions are always satisfied, since $%
a=0.81,\ 0.1904<b<0.2875,\ 0.8733<d<0.9980.$
\end{pf}

\FRAME{ftbpFU}{4.4417in}{3.3347in}{0pt}{\Qcb{The map $g$, a typical
irregular (and chaotic) orbit and the bifurcaion diagram when $\protect%
\alpha $ is varied}}{\Qlb{fig1}}{mendesdfig1.png}{\special{language
"Scientific Word";type "GRAPHIC";maintain-aspect-ratio TRUE;display
"USEDEF";valid_file "F";width 4.4417in;height 3.3347in;depth
0pt;original-width 8.0073in;original-height 6.0001in;cropleft "0";croptop
"1";cropright "1";cropbottom "0";filename
'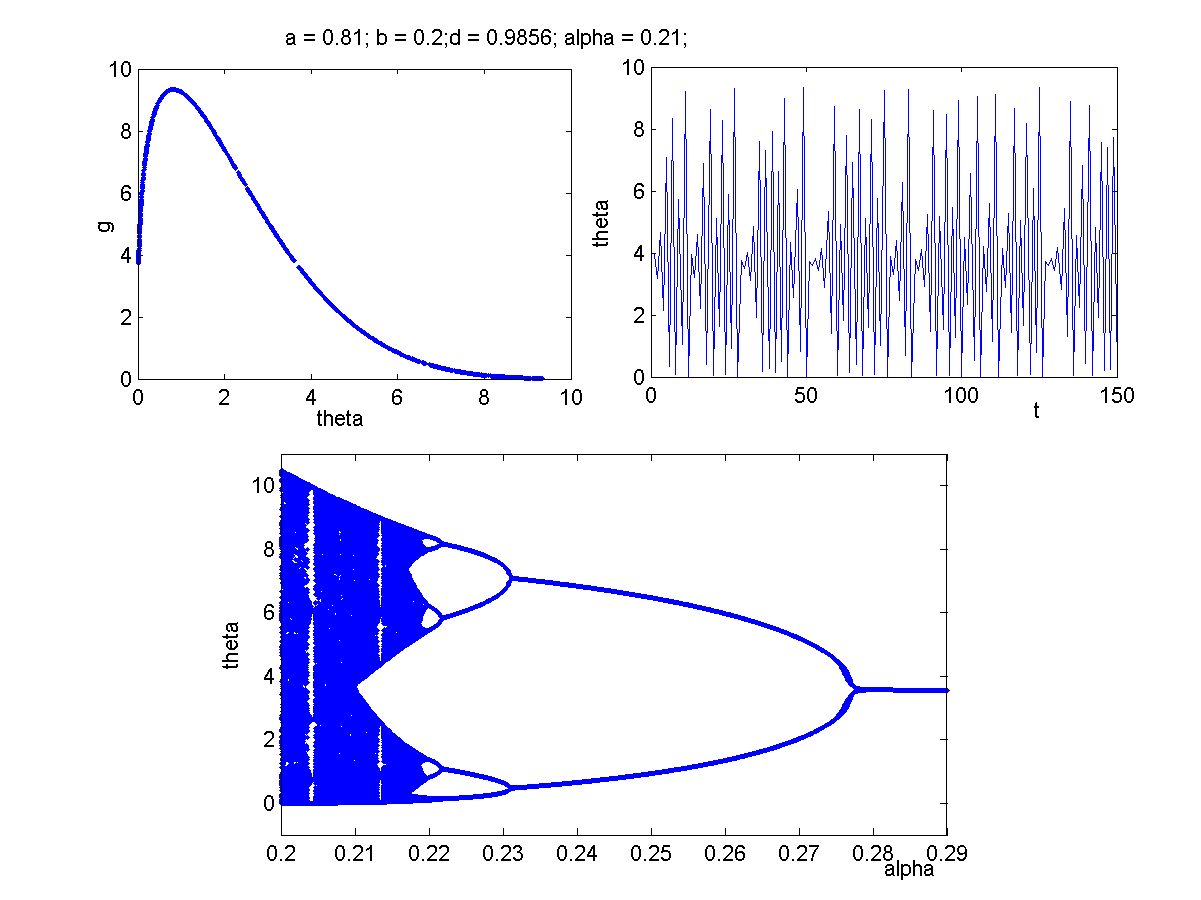';file-properties "XNPEU";}}

Moreover, in Figure \ref{fig2} the bifurcation diagram is illustrated when
the parameter $a$ is varied between $0.7$ and $0.9$ and taking the same
parameter calibration as in the above Proposition: $\alpha =0.21\ ,b=0.2$
and $d=0.9868.$ There is also a period-doubling route to chaos, but the map $%
g$ dynamics escapes to infinity before the occurrence of a period three
orbit. This presents one more reason to consider the complete Sharkowski
relation between the periodic orbits of the map and to apply some other
techniques in the search for chaos in this specific model. In the same
figure we also represent the variation of the Lyapunov exponent for $a$
belonging to the interval $[0.7,0.9].$ We recall that positive Lyapunov
exponent is a necessary condition for chaos.

\FRAME{ftbpFU}{4.4417in}{3.3347in}{0pt}{\Qcb{Bifurcation diagram when the
parameter $a$ is varied and corresponding Lyapunov exponent.}}{\Qlb{fig2}}{%
mendesdfig2.png}{\special{language "Scientific Word";type
"GRAPHIC";maintain-aspect-ratio TRUE;display "USEDEF";valid_file "F";width
4.4417in;height 3.3347in;depth 0pt;original-width 8.0073in;original-height
6.0001in;cropleft "0";croptop "1";cropright "1";cropbottom "0";filename
'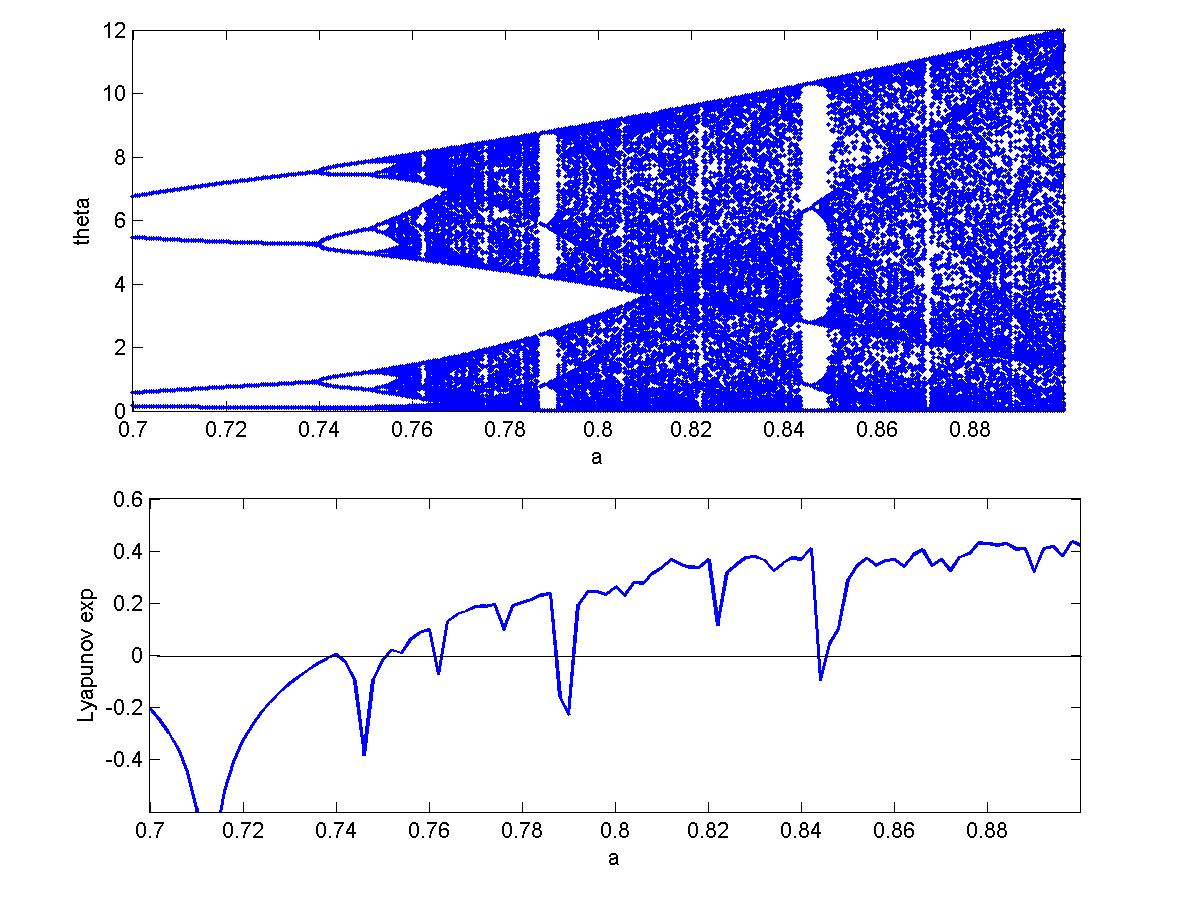';file-properties "XNPEU";}}

Bhattacharya and Bunzel \cite{BB} suggest three examples for the study of
the dynamics of the map $g$. In the first case a period 3-cycle is found,
which implies the existence of chaos in the Li-Yorke sense if the Sharkovsky
order is applied. In the second example, it was argued that a period 3-orbit
could not be found for the following parameter values: $a=0.75,$ $b=0.58,$ $%
d=0.62,$ $\alpha =0.15$ since the equation $g^{3}\left( \theta \right)
=\theta $ has no solution. Since the map $g$ is unimodal and Mitra's
sufficient condition for chaos in unimodal maps is verified, the conclusion
was that for this parameter setting chaotic motion in the backward dynamics
could also be found.\footnote{%
As argued in \cite{Mit}, for a continuous unimodal map $f:X\rightarrow X,$
where $X$ is a non-negative interval, $x_{\max }$ is the critical point such
that $f\left( x_{\max }\right) >x_{\max }$ and $x_{\ast }$ is the unique
fixed point of the map such that $x_{\ast }>x_{\max },$ Mitra states that:
If $f$ satisfies $f^{2}\left( x_{\max }\right) <x_{\max }$ and $f^{3}\left(
x_{\max }\right) <x_{\ast }$, then $\left( X,f\right) $ shows topological
chaos.} Finally, in the third example the set of parameters are: $a=0.9,$ $%
b=0.7,$ $d=0.6,$ $\alpha =0.2.$ For these values no period three orbit is
found, neither the sufficient condition of Mitra is verified. Therefore, it
was argued that for this case the very existence of chaos for the unimodal
map is questioned on the grounds of a lack of logical proof of such dynamics.

In order to clarify some of the issues raised by \cite{BB}, a symbolic
dynamics approach is developed for the unimodal map $g$, which allow us to
perform the computation of the topological entropy for any choice of
parameters, and, of course, permit us to classify the complexity of the map
since positive topological entropy implies the existence of chaotic
dynamics. We will concentrate on their third example.

We consider again the unimodal map $g:\left[ 0,g_{\max }\right] \rightarrow %
\left[ 0,g_{\max }\right] .$ This kind of map has symbolic dynamics relative
to a topological Markov partition generated by the orbit of the critical
point $\theta _{\max }.$ This is illustrated in Figure \ref{FigMarkov} for
the parameter values presented in Example 2. So, any numerical trajectory $%
\theta _{0}=\theta _{\max },\theta _{1},\theta _{2},...$ for the map $g$
corresponds to a symbolic sequence $\sigma _{0}\sigma _{1}\sigma _{2}...$
where $\sigma _{i}\in \{L,C,R\}$ depending on where the point $\theta _{i}$
falls in, i.e.,%
\begin{equation*}
\sigma _{i}\left( \theta _{0}\right) =\left\{ 
\begin{array}{cc}
L & \text{if }g^{i}\left( \theta _{0}\right) <\theta _{\max } \\ 
C & \text{if }g^{i}\left( \theta _{0}\right) =\theta _{\max } \\ 
R & \text{if }g^{i}\left( \theta _{0}\right) >\theta _{\max }%
\end{array}%
\right. .
\end{equation*}%
All symbolic sequences made of these letters may be ordered by the natural
lexicographical order $L<C<R$ and

\FRAME{ftbpFU}{4.1295in}{3.5561in}{0pt}{\Qcb{Markov partition for a period 5
orbit.}}{\Qlb{FigMarkov}}{mendesdfig3.eps}{\special{language "Scientific
Word";type "GRAPHIC";maintain-aspect-ratio TRUE;display "USEDEF";valid_file
"F";width 4.1295in;height 3.5561in;depth 0pt;original-width
6.5959in;original-height 5.6775in;cropleft "0";croptop "1";cropright
"1";cropbottom "0";filename '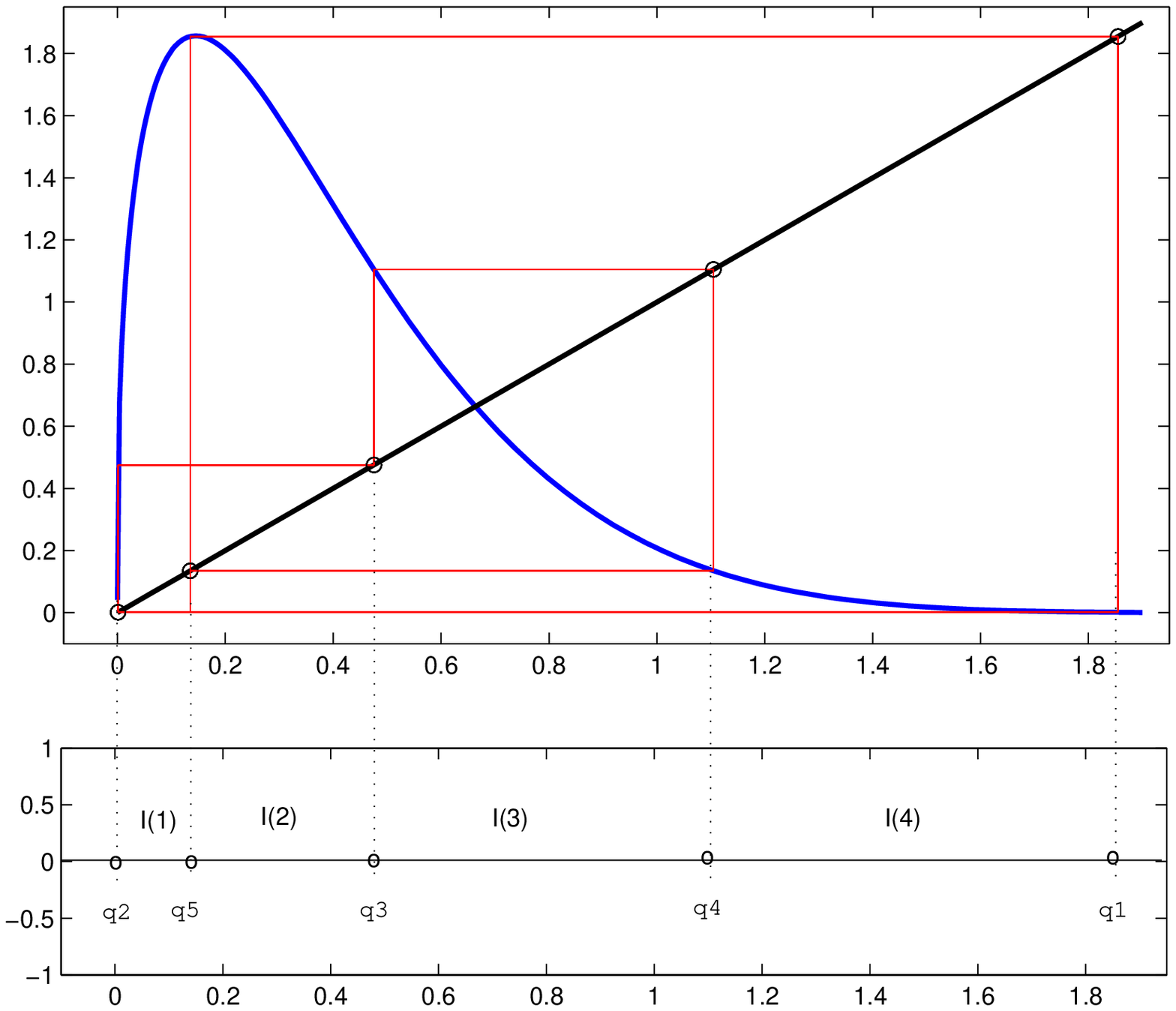';file-properties "XNPEU";}}

Defining the fullshift $\Sigma _{2}=\left\{ \sigma =\sigma _{0}\sigma
_{1}\sigma _{2}....\text{ where }\sigma _{i}=L\text{ or }R\right\} $ to be
the set of all possible infinite symbolic strings of $L$'s and $R$'s, then
any given infinite symbolic sequence is a singleton in the fullshift space.
The Bernoulli shift map $s:\Sigma _{2}\rightarrow \Sigma _{2}$ is defined by 
$s\left( \sigma \right) =s\left( \sigma _{0}\sigma _{1}\sigma _{2}...\right)
=\sigma _{1}\sigma _{2}\sigma _{3}...$. In general, not all symbolic
sequences correspond to the trajectory of an initial condition $\theta _{0}.$
Restricting the shift map to a subset of $\Sigma _{2}$ consisting of all the
itineraries that are realizable yields the subshift $\Sigma \subset \Sigma
_{2}.$

We formulate the result in terms of topological Markov chains, a special
class of subshifts of finite type where the transition in the symbol
sequence is specified by a $0-1$ matrix. Any $\left( n\times n\right) $
binary matrix $M=\left( M_{ij}\right) _{i,j=0,...,n-1},\;M_{ij}\in \left\{
0,1\right\} $ generates a special subshift 
\begin{equation*}
\Sigma _{M}=\left\{ \sigma \in \Sigma _{2}:M_{\sigma _{i}\sigma
_{i+1}}=1,\;\forall i\in \mathbb{N}\right\}
\end{equation*}%
which is called the topological Markov chain associated with the Markov
matrix $M$. We say that $M_{\sigma _{i}\sigma _{i+1}}=1$ if the transition
from $\sigma _{i}$ to $\sigma _{i+1}$ is possible. The matrix $M$ gives a
complete description of the dynamics of the unimodal map. The premier
numerical invariant of a dynamic system is its topological entropy defined
by 
\begin{equation*}
h_{top}\left( \Sigma _{M}\right) =\lim_{n\rightarrow \infty }\frac{\log
\left( \sharp W_{n}\left( \Sigma _{M}\right) \right) }{n},
\end{equation*}%
where $W_{n}\left( \Sigma _{M}\right) $ is the set of words of length $n$
occurring in sequences of $\Sigma _{M}$. Moreover, if the topological Markov
matrix $M$ is given then, the topological entropy is the natural logarithm
of the spectral radius of $M$.

For the parameter values $a=0.75,$ $b=0.58,$ $d=0.62,$ $\alpha =0.15$, we
found a period 5 orbit: $\left\{ 1.8549,0.0013,0.4756,1.1047,0.1350\right\} $
which is shown in Figure \ref{FigMarkov} with the corresponding 4 interval $%
\left\{ I_{i}\right\} _{i=1,...,4}$ Markov partition. The critical point
assumes the value $\theta _{\max }=0.1452$ and generates the symbolic
partition for the map $g.$ The periodic orbit has the following symbolic
address: $\left( \theta _{1}\theta _{2}\theta _{3}\theta _{4}\theta
_{5}\right) ^{\infty }=\left( RLRRC\right) ^{\infty }$ and in consequence we
have the following Markov matrix:%
\begin{equation*}
M_{RLRRC}=\left[ 
\begin{array}{cccc}
0\; & 0\; & 1\; & 1 \\ 
0\; & 0\; & 0\; & 1 \\ 
0\; & 1\; & 1\; & 0 \\ 
1\; & 0\; & 0\; & 0%
\end{array}%
\right] .
\end{equation*}%
The maximal eigenvalue of $M$ is given by $\lambda =1.5128,$ which implies
that the topological entropy is positive: $h_{top}\simeq 0.4140$ and this
shows very clearly that we are dealing with chaotic motion in this set of
parameter values.

It should be noted that for any other kneading sequence, for a suitable
choice of parameters values, we can obtain a Markov partition and a Markov
matrix which totally determine the complexity of the unimodal map. This is a
very simple and rigorous way to estimate the topological entropy of a
one-dimensional model and to check for the existence of chaos.

\section{Concluding Remarks}

The general use of the Li-Yorke theorem or the Mitra condition may lead to
misleading conclusions about the existence of chaotic dynamics, as it was
done in \cite{BB}. Therefore, in order to obtain relevant answers to whether
there are or not chaotic dynamics under certain ranges of parameters values
in a 1-dimensional particular model, we suggest that a bifurcation diagram,
the variation of the Lyapunov exponent, the existence of a periodic point of
period not equal to a power of two, and symbolic dynamics are very powerful
techniques for that purpose. Moreover, the application of these techniques
to the matching labor market model so interestingly developed by
Bhattacharya and Bunzel clearly confirmed that a very simple model of the
labor market, with well behaved aggregate functions (continuous, twice
differentiable and linearly homogeneous) do really produce chaotic behavior
for a range of parameter sets which had been questioned in \cite{BB1}, for
high values of the measure of labor market tightness.

Financial support from the Funda\c{c}\~{a}o Ci\^{e}ncia e Tecnologia,
Lisbon, is grateful acknowledged, under the contract No POCTI/ ECO /48628/
2002, partially funded by the European Regional Development Fund (ERDF).

\end{document}